\documentclass{article}
\usepackage{setspace}
\usepackage{orcidlink}
\usepackage{hyperref}
\usepackage{titlesec}
\usepackage{authblk}
\usepackage{makecell}
\usepackage{adjustbox}
\usepackage{graphicx}
\usepackage[normalem]{ulem}
\usepackage{multirow}
\usepackage{amsmath}
\usepackage[style=numeric, backend=biber]{biblatex}

\date{}

\begin{document}

\title{Quantifying Institutional Gender Inequality in Contemporary Visual Art}

\author[a]{\orcidlink{0009-0005-1288-2076} Xindi Wang}
\author[b]{\orcidlink{0000-0003-0099-7480} Alexander J. Gates} 
\author[c,d,e]{\orcidlink{0000-0002-6442-8415} Magnus Resch}
\author[a,f,g,h,*]{\\ \orcidlink{0000-0002-4028-3522} Albert-L\'{a}szl\'{o} Barab\'{a}si}

\affil[a]{Network Science Institute, Northeastern University, Boston, MA, USA}
\affil[b]{School of Data Science, University of Virginia, Charlottesville, VA, USA}
\affil[c]{Yale School of Management, Yale University, New Haven, CT, USA}
\affil[d]{Zagreb School of Economics and Management, Zagreb, Croatia}
\affil[e]{Luxembourg School of Business, Luxembourg}
\affil[f]{Department of Mathematics and its Applications and Center for Network Science, Central European University, Budapest, Hungary}
\affil[g]{Department of Network and Data Science, Central European University, Budapest, Hungary}
\affil[h]{Division of Network Medicine, Department of Medicine, Harvard Medical School, Boston, MA, USA}
\affil[*]{To whom correspondence should be addressed: \href{mailto:a.barabasi@northeastern.edu }{a.barabasi@northeastern.edu}}

\maketitle
\vspace{-0.5in}
\begin{abstract}
    From disparities in the number of exhibiting artists to auction opportunities, there is evidence of women's under-representation in visual art.
    Here we explore the exhibition history and auction sales of 65,768 contemporary artists in 20,389 institutions, revealing gender differences in the artist population, exhibitions and auctions.
    We distinguish between two criteria for gender equity: gender-neutrality, when artists have gender-independent access to exhibition opportunities, and gender-balanced, that strives for gender parity in representation, finding that 58\% of institutions are gender-neutral but only 24\% are gender-balanced, and that the fraction of {man-overrepresented} institutions increases with institutional prestige.
    We define artist's co-exhibition gender to capture the gender inequality of the institutions that an artist exhibits. Finally, we use logistic regression to predict an artist's access to the auction market, finding that co-exhibition gender has a stronger correlation with success than the artist's gender.
    These results help unveil and quantify the institutional forces that relate to the persistent gender imbalance in the art world.
\end{abstract}

\pagebreak
\setstretch{1.5}

\section*{Introduction}
Linda Nochlin's question, Why have there been no great women artists?~\cite{nochlin1971have}, formulated fifty years ago, today is informed by widespread awareness of gender disparities across all disciplines, from science to industry. 
It is also supported by evidence of women's under-representation in the art world, captured by multiple metrics, from disparities in the number of women and men artists~\cite{davis2016why, galenson2007were, lozano2019visual, greenwald2021painting}, to the lack of gender balance in museums and galleries~\cite{reilly2019taking, lozano2019visual, halperin2019museums, halperin2019methodology, topaz2019diversity}, and the lack of women artists in auction sales~\cite{adams2021gendered, halperin2019female, bocart2022empirical, cameron2019art}.
The root of these gender differences remains a topic of debate, with hypotheses ranging from genetic predisposition to discrimination~\cite{cowen1996women, adams2021gendered, leblanc2022women}.
In this article, we focus on the potential relationship of institutional practices on gender inequality in the art world. 
Regardless of whether institutional practices are grounded in potential differences in the artwork itself, or reflect discrimination of the artists, ultimately institutional practices combine with reputation shape access to representation, resources and rewards in art~\cite{fraiberger2018quantifying, mitali2018fame,elgammal2018shape}.
These mechanisms are exacerbated by lock-in effects that limit the access of some artists to future opportunities, while offering winners-take-all outcomes for others~\cite{fraiberger2018quantifying}.
The interlocking effects suggest that to truly understand gender inequality in art, we need to understand the ecosystem that surrounds the production and the consumption of art, from the complete exhibition careers of all artists to the exhibition record of all institutions, together with the auction record of the secondary market, as they together shape the structural and institutional patterns that are related to gender bias. 
This research is methodologically aided by recent advances in quantifying the evolution, roots, and impact of gender disparities in science and other areas~\cite{lariviere2013bibliometrics, lincoln2012matilda, holman2018gender, zippel2017women, huang2020gender, etzkowitz2000athena, fortunato2018science, nakandala2017gendered, dworkin2020extent, nettasinghe2021emergence, teich2022citation, lerman2022gendered}.
As scientific discourse is increasingly rooted in data, its production and impact is catalogued in a fashion that makes it amenable for quantitative analysis. 
In contrast, systematic and reliable datasets that allow us to quantify artistic careers have been difficult to develop and explore.

Here, we remedy this situation by exploring a dataset with a broad coverage, capturing the exhibition careers of 496,354 artists in 23,570 institutions, representing both museums and galleries~\cite{fraiberger2018quantifying, sinatra2016quantifying}.
For the purposes of this study, we restricted our focus to contemporary, active artists who began exhibiting on and after 1990, hence their exhibition history can be accurately reconstructed, resulting in 65,768 artists participating in 378,517 exhibitions at 20,389 institutions worldwide (Fig.~\ref{fig:fig1}a, b, c). 
This focus is motivated by our goal of identifying relationships between institutional gender representation and active artist careers.
To quantify gender-related patterns, we combine expert curated gender information with a gender identification service based on artists' names \href{https://genderize.io/}{{genderize.io}}(\cite{shen2013mind, holman2018gender}), allowing us to assign most artists a binary gender (See Methods section).
To understand gender inequality of each art institution, we inspect each institution under two criteria of gender equity: gender-neutrality, when artists have gender-independent access to exhibition, and gender-balance, which aims for gender parity in representation. 
We find that 58\% of art institutions are gender-neutral but only 24\% are gender-balanced, and the fraction of man-overrepresented institutions increases with institutional prestige. 
To connect institutional gender inequality with an artist's career, we define the co-exhibition gender of an artist to capture the gender inequality of the institutions that an artist exhibits under gender-neutral criteria. 
We find that artists with high prestige are more likely to be labelled as co-exhibition man, and co-exhibition gender tends to remain stable over an artist's career. 
Finally, we build a logistic regression model to predict an artist's access to the auction market, finding that co-exhibition gender has a stronger association with access to the auction market than artist's gender.

\section*{Results}
\subsection*{Gender Inequality in Artist Population and Access to Exhibition}
To begin, we identify gender differences in the number of artists and exhibitions.
This initial overview provides context for the more nuanced analysis which follows and empowers us to capture institutional standpoint of gender imbalances in the art world.
Our sample of 65,768 artists contains 41,738 men and 24,030 women, which corresponds to 1.74 men artists for every women artist (Fig.~\ref{fig:fig1}a, Table~\ref{tab:bias_table}). 
This ratio captures the inherent population disparity, an elementary and widely explored evidence of gender imbalance~\cite{davis2016why, halperin2019methodology, lozano2019visual}.
However, population disparity reflects only one dimension of gender inequality. 
Another important metric is the exhibition disparity, which focuses on an artist's access to exhibition opportunities.
We found that men artists were exhibited 662,571 times, in contrast to women artists who were exhibited only 355,506 times, resulting in an exhibition gender ratio of 1.86 (Fig.~\ref{fig:fig1}a, Table~\ref{tab:bias_table}).
This exhibition disparity is higher than the population disparity, indicating not only that men artists outnumber women artists, but that men artists get more access to exhibition opportunities than expected based on their presence in the artist population.

Taken together, these results support previous observations relying on fragmented or local data, used by art theorists, activists, and artists to urge institutions to acknowledge and address gender imbalance~\cite{halperin2019museums, halperin2019methodology, lozano2019visual, reilly2019taking, halperin2019female}. 
These data further document a persistent gender imbalance both in terms of the number of exhibiting artists, and in terms of access to exhibition and auction venues. 

\subsection*{Gender-balance and Gender-neutrality in Art}
Museums and galleries today are under tremendous professional, community, and societal pressure to address the widespread gender imbalance present in the art world.
Operationally, however, this conversation becomes muddled, since there are many different criteria for assessing gender equality in art.
Here, we quantitatively explore our progress under two criteria: 1)
 gender-neutral hypothesis, i.e., the expectation that all artists, independent of gender, should have the same equitable access to exhibition opportunities (Table~\ref{tab:def}); 
and 2) gender-balance, i.e., the important but different goal of achieving a 50/50 gender ratio in both population and representation in art (Table~\ref{tab:def}).
Critically, gender-balance and gender-neutrality are two different objectives, which require different metrics and different kinds of interventions.
Gender-balance, which aims for gender parity (50/50) in the artist population and representation, is a primary goal towards which the art community should and does strive.
Achieving it requires a strategy with a longer time horizon, including investment in art education, balanced or affirmative enrolment in art programs, and most importantly, career-nurturing programs that help reduce the dropout of women artists, the root of gender-imbalance in many professions~\cite{huang2020gender}.
In contrast, gender-neutrality represents a more immediate goal of offering equal opportunities to all currently exhibiting artists, independent of their gender. 
Hence, gender-neutral representation could be achieved over shorter timescales, by engaging with the current gatekeepers of access, like curators, collectors, and directors at museums and galleries, offering them the metrics and structural insights to empower them to offer balanced representation.

Ultimately, the two goals are intertwined: to achieve a sustainable 50/50 gender ratio in exhibitions and auctions (gender-balance), it's not sufficient to focus only on the artist population, but we must correct the institutional forces that affect gender-neutrality and inhibit equal exhibition and auction access to women and men.
While in recent years, enrolment in many art schools has achieved gender parity~\cite{cameron2019art, NMWA2021get}, the representation of those graduates is not 50/50, indicating a lack of institutional gender-neutrality in the art world.
Yet, a gender-neutral objective alone has its limitations: it freezes the current status quo, and while it may improve access to exhibition opportunities for the currently practising women artists, it is only a step towards, but not a guaranteed path to gender-balance. 
In the following, our goal is to independently explore gender equality in art in terms of both gender-balance and gender-neutrality, asking how well the institutions that support the art world do in terms of these two criteria.

\subsection*{Institutional Representation Inequality}
Inequalities in the institutional representation of artists are revealed by the extent to which the gender ratio of an institution's list of exhibitions deviates from the baseline composition of the artist population.
As only 36.54\% of contemporary exhibiting artists are women (Figure~\ref{fig:fig1}a, Table~\ref{tab:bias_table}), the active artist population is highly gender-imbalanced.
A gender-neutral institution, which offers equal access to all exhibiting artists, must therefore devote 36.54\% of their exhibitions to women artists. 
In contrast, a gender-balanced institution strives for a 50/50 representation, independent of the gender composition of the artist population. 
It is important to clarify that the definitions of gender-neutrality and gender-balance are based on statistical observations of outcomes, and are not meant to infer the factors used by institutions in their decision-making practice when selecting artists for exhibition.

Both gender-balance and gender-neutrality can be mathematically conceptualized with a two-tailed binomial hypothesis test, measuring the deviation between the observed proportion $p_i$ of exhibitions by women artists in institution $i$ and the null hypothesis of $p_0$. 
In both cases, an institution is assumed to be independently drawing their exhibitions from a universal pool of artists with replacement.
Gender-balance implies $p_0 = 0.5$ (50\% of the exhibitions are of women artists), while the gender-neutral goal corresponds to $p_0=0.365$ (corresponding to the current fraction of 36.54\% contemporary women artists).
Using the recorded number of exhibitions $n_i$ as the sample size for each institution, the binomial test's Bayes Factor (BF) categorizes an institution $i$ into one of four possible categories (further details in Methods section): 
institutions in which men are overrepresented (man-overrepresented; $p_i < p_0$) or institutions in which women are overrepresented (woman-overrepresented; $p_i > p_0$), whose proportion of exhibitions featuring women artists is statistically smaller or larger than the null hypothesis;
gender-neutral or gender-balanced institutions, whose proportion of women and men artists is statistically indistinguishable from the null hypothesis ($p_i \approx p_0$); and 
uncategorised institutions, that lack sufficient evidence for categorization as indicated by their Bayes Factors.
We find that under the gender-balanced criteria, 12,489 institutions are uncategorised while under gender-neutral criteria, 11,493 institutions are uncategorised. Unless noted otherwise, we ignore these uncategorised institutions in our analysis. 
However, due to the heavy-tailed nature of the exhibition distribution~\cite{fraiberger2018quantifying}, the uncategorised institutions account for only27.64\% and 19.85\% of all exhibitions under the gender-balanced and gender-neutral criteria respectively, and therefore represent a minor contribution to most artists' careers. 

We find that 57.80\% of the art institutions are gender-neutral, but only 24.46\% of the art institutions are gender-balanced (Fig.~\ref{fig:fig2}a).
Under the gender-neutral criteria, 23.58\% institutions are man-overrepresented and 17.62\% institutions are woman-overrepresented(Fig.~\ref{fig:fig2}a). 
In contrast, under the gender-balanced criteria, 72.19\% institutions are man-overrepresented and only 3.35\% institutions are woman-overrepresented (Fig.~\ref{fig:fig2}a).

Under the gender-neutral criteria (Fig.~\ref{fig:fig2}a), we find that a higher fraction of museums than galleries are gender-neutral (69.10\% of museums vs 51.60\% of galleries).
Moreover, galleries show a stronger positive bias towards men artists (1541 man-overrepresented vs 929 woman-overrepresented), while museums are split between man-overrepresented (401) and woman-overrepresented (463).
Under the gender-balanced criteria, we again observe that a higher fraction of museums are gender-balanced (27.77\% of museums vs 22.76\% of galleries), however we find a higher fraction of galleries that embrace women artists than museums (3.76\% of galleries vs 2.55\% museums), indicating the emergence of a small number of galleries that nurture the career of women artists.

Figure \ref{fig:fig2}d shows the number of men and women exhibitions for 100 prominent institutions worldwide, plotted alongside the decision boundary for the gender-neutral and gender-balanced criteria based on the binomial test.
The figure indicates, for example, that MoMA (The Museum of Modern Art) is gender-neutral (BF: $0.108$) but man-overrepresented (BF: $4.229 \times 10^7$) under the gender-balanced hypothesis; Whitney Museum of American Art is woman-overrepresented (BF: $4.518$) under the gender-neutral criteria but man-overrepresented (BF: $10.697$) under gender-balanced criteria; and the Albertina is woman-overrepresented under both gender-neutral (BF: 45828.026) and gender-balanced (BF: $4.400$) criteria.
A similar classification for countries/regions (Fig.~\ref{fig:fig3}b, c) reveals that most countries/regions are man-overrepresented under the gender-balanced criteria.
However, under the gender-neutral criteria (Fig.~\ref{fig:fig3}a, c), we can identify woman-overrepresented exhibition patterns in Austria (BF: $10^{73}$), Sweden (BF: $3.660 \times 10^{64}$), Finland (BF: $6.183 \times 10^{62}$), etc., and we find that United States (BF: $4.344 \times 10^{-12}$), Germany (BF: $2.353 \times 10^{-17}$), United Kingdom (BF: $0.027$) etc. are characterized by gender-neutral exhibition patterns.

We next consider how institutional representation inequality varies with the institutional prestige. 
In the visual arts, reputation and networks of influence play an important role in determining access to resources and rewards.
Here, we adopt a network-based measure of institutional prestige derived from the artist co-exhibition network, a directed weighted network connecting two institutions if an artist exhibits in the source institution before exhibiting in the target institution~\cite{fraiberger2018quantifying}.
Specifically, the eigenvector centrality of an institution in the co-exhibition network correlates strongly with both expert assessment of institutional prominence and the average auction price of the exhibited artists, empowering us to assign a prestige score to all art institutions in our dataset~\cite{fraiberger2018quantifying}.
To reveal the relationship between institutional prestige and gender representation inequality, we group institutions into three prestige categories: low prestige (lower than 40th percentile of all institutions), mid prestige (between 40th and 70th percentile) and high prestige (higher than 70th percentile). 
We find that under the gender-neutral hypothesis, the fraction of man-overrepresented institutions increases with the institutional prestige, while the fraction of {woman-overrepresented} institutions decreases (Fig.~\ref{fig:fig2}b). 
The effect is particularly strong under the gender-balanced hypothesis: while only 67.06\% of the low prestige institutions are man-overrepresented, 80.20\% of the high prestige institution are in this category (Fig.~\ref{fig:fig2}c). 
We also find that the percentage of woman-overrepresented institutions is tiny (between 1.41\%-5.30\%) in all three prestige categories (Fig.~\ref{fig:fig2}c).
These results are robust to the specific number and placement of the prestige bins, and are qualitatively replicated on the subset of institutions with expert prestige scores (Supplementary Information S2.5, Supplementary Figure 5).

To offer an overview of the distribution of institutional inequality across the whole institutional space, in Fig.~\ref{fig:viz}a we show the art institution network~\cite{fraiberger2018quantifying}, whose nodes are the museums and galleries, connected to each other if they exhibit the same artists in a statistically significant number. 
The network is laid out using a force-directed algorithm, that tends to place close to each other directly linked institutions, and helps uncover densely connected communities. 
We coloured each node to capture the respective institution's gender inequality status: man-overrepresented (blue), woman-overrepresented (red), and gender-neutral (gray). 
The emergence of clusters dense in blue or red nodes is evidence of inequality-based assortativity, indicating that institutions with comparable gender representation tend to be connected to each other, often exhibiting the same artists. 

To better illustrate this, we also show a case study on the sub-graphs of several European countries: Germany, Austria, France, Spain, and Portugal (Fig.~\ref{fig:viz}b).
The sub-graph illustrates the regional differences: Germany and Austria have more woman-overrepresented institutions while France, Spain, and Portugal have more man-overrepresented institutions. 
It also captures clusters based on gender representation within a region: for example, we can see that woman-overrepresented institutions in Austria are close to woman-overrepresented institutions in Germany, while the few man-overrepresented institutions in Austria are more close to man-overrepresented institutions in other countries.
To quantify the magnitude and significance of this gender-representation assortativity, we calculate the percentage of weighted outgoing links to institutions with each gender representation under the gender-neutral criteria. 
We find that man-overrepresented institutions connect to other man-overrepresented institutions more than expected by institutions randomly connected to each other, 43.7\% compared to the random baseline of 23.58\%. 
Similarly, woman-overrepresented institutions connect to other woman-overrepresented institutions more than would be expected by institutions randomly connected to each other, 25.1\% compared to the random baseline of 17.62\% (Supplementary Information S1.2).
The institutional variability in assortativity can be captured using a measure of multi-scale assortativity~\cite{peel_multiscale_2018} (Supplementary Information S1.2), which further illustrates the extent to which gender-representation shapes the connectivity of the co-exhibition network.

In summary, we find that institutions vary widely in terms of their ability to establish a gender-neutral or gender-balanced exhibition schedule.
Among categorisable institutions, 57.80\% are gender-neutral, embracing a list of exhibitions that mirror the gender mix of the artist population.
The institutions fare worse in terms of gender-balance, the majority of institutions displaying man-overrepresented exhibition patterns.
Overall, museums are more likely to be both gender-neutral and gender-balanced than galleries. 
At the same time, we do observe the emergence of a small number of galleries that strongly embrace and promote women artists.
Finally, we find that the proportion of institutions characterized by gender-balance and gender-neutrality varies with the prestige of the institutions: the higher the prestige, the institutions' exhibition patterns are more biased towards men artists.

\subsection*{Co-exhibition Gender of Artists}
We next explore the relationship between institutional gender representation inequality and artists' careers.
For this, we assign to each artist with more than 10 exhibitions a co-exhibition gender under the gender-neutral criteria for institutional gender representation (see Supplementary Information S2.2 for a similar analysis under gender-balanced criteria), to reflect the dominant co-exhibition community an artist exhibited in: an artist is labelled as co-exhibition woman (co-exhibition man/co-exhibition neutral) if the artist's exhibition history contains an over-representation of woman-overrepresented (man-overrepresented/gender-neutral) institutions compared to what we would have expected if the exhibition institutions were selected at random (formal definition in Methods section). 
For 47.8\% of men artists and 49.3\% of women artists, the co-exhibition gender agrees with the artist's gender.
At the same time, 19.6\% of women artists are labelled as co-exhibition man, 19.7\% of men artists are labelled as co-exhibition woman, and about 32\% of all artists from both genders are labelled as co-exhibition neutral (Fig.~\ref{fig:fig5}a).
If we consider the relationship with an artist's individual prestige (calculated as the average prestige of the institutions that exhibited the artist), both man and woman artists with high prestige are less likely to be labelled as co-exhibition woman (Fig.~\ref{fig:fig5}b).

How stable is an artist's co-exhibition gender designation, i.e., does it change throughout an artist's career? 
For this, we compare the co-exhibition gender of an artist's first 5 exhibitions to the co-exhibition gender of the artist's last 5 exhibitions.
Figure~\ref{fig:fig5}c shows the fraction of artists labelled using their early-career exhibitions (rows) and then using their late-career exhibitions (columns), both under the gender-neutral criteria (same analysis on gender-balanced criteria see Supplementary Information S2.2). 
We find that the highest density lies on the matrix diagonal, indicating that for all three categories (co-exhibition man, co-exhibition woman, and co-exhibition neutral) there is co-exhibition gender lock-in. 
In other words, the early co-exhibition gender label of an artist tends to match the late co-exhibition gender label.
Overall, co-exhibition man artists have the highest lock-in intensity: 73\% of initially co-exhibition men artists remain co-exhibition men, while only 54\% of co-exhibition women artists and co-exhibition neutral artists maintain their initial label. 
The strength of this co-exhibition gender lock-in decreases slightly with an artists' career prestige (Fig.~\ref{fig:fig5}d), yet even at high prestige only 11\% of co-exhibition man artists transition to co-exhibition woman.
Furthermore, the percentage of co-exhibition men artists that change their co-exhibition gender is less than co-exhibition women and co-exhibition neutral artists, indicating that the path towards high prestige institutions goes through man-overrepresented institutions.

\subsection*{Gender Disparity in Auctions}

Sales, both in the primary (galleries) and the secondary (auctions) market, and the resulting valuation of an artist's work, is a frequently used measure of artistic success.
While gender disparity has been extensively documented in auction sales~\cite{allen2005xfactor, adams2021gendered, leblanc2022women, bocart2022empirical}, it is unclear if its degree reflects the gender disparities already present in the artist population and exhibition patterns, or represents additional dimensions of gender disparity.
To address this question, we linked the exhibition record of each artist to his/her auction records (if any), allowing us to explore the degree to which gender disparities in the artist population and exhibition patterns translate to gender disparity in auction sales.

We find that 10,179 men and 3,761 women artists recorded at least one sale at auction (Fig.~\ref{fig:fig6}a, Table~\ref{tab:bias_table}), revealing an auction population disparity of 2.71, larger than the population disparity (1.74) and the exhibition disparity (1.86).
The gender gap is even more pronounced when we inspect the number of auctioned items (records): of the 125,682 auction records, 81.74\% are for art created by men artists (102,729), reflecting an auction record disparity of 4.48 (Fig.~\ref{fig:fig6}a, Table~\ref{tab:bias_table}). 
Looking at total auction sales in volume (calculated using normalized price), we found an auction sales disparity of 7.54 (Fig.~\ref{fig:fig6}a, Table~\ref{tab:bias_table}).

Combining exhibition and auction records allows us to explore the rate at which artists transition from exhibitions to auctions.
We find that 21.83\% of the artists that exhibit have at least one recorded auction sale, an aggregate rate that favours men artists: 24.39\% of the exhibiting men artists transitioned to auction, in contrast with only 15.65\% women artists. 
The number of auctions and the auction prices are also biased towards men artists (Table~\ref{tab:bias_table}): on average men artists have 1.66 times more auctions than women artists (10.1 vs 6.1) and command a 1.75 times higher auction price (0.7 vs 0.4, normalized).
Taken together, these numbers indicate that the auction process does not simply mirror the observed gender imbalance in terms of the artist population or exhibition patterns, but exacerbates that, representing a new dimension of gender disparity.

\subsection*{The Relationship Between Institutional Representation Inequality and Auctions}

Most importantly, the auction market dynamics allow us to quantify the relative importance of the different variables that correlate with the access to the auction market, helping us to gain a deeper understanding of gender and co-exhibition representation inequality. 
To achieve this, we built a series of logistic regression models (see Table \ref{tab:lr}) to predict the probability $P(a_i=1)$ that artist $i$ transitions to the auction market,
first given only his/her average exhibition count per year, and career length in Model 1.
Then, in Model 2, we add the artists' gender, and in Model 3 the co-exhibition gender under gender-neutral criteria. 
For the gender and institutional gender variables, we employed a dummy-encoding with a baseline value of man and man-overrepresented respectively.
Finally, in Model 4, we combine the artists' gender and co-exhibition gender through an interaction term. Details of the logistic regression model can be found in Methods.

Rather than predictive ability, our focus here is on the relative importance of each variable measured by the odds ratio.

The model suggests a notable association between the number of exhibitions per year and career length with the likelihood of entering the auction market, showing that artists that are frequently exhibited and/or have a long career are more likely to enter the auction market, which is in line with the trends seen in Fig.~\ref{fig:fig6}b, c.

The addition of artist gender into Model 2 reduces the Bayesian Information Criterion (BIC), a general measure of goodness of fit based on the log-likelihood discounted by the number of parameters used in the model, indicating a model improvement.
The artist's gender has a modest effect on auction access and reveals the different roles gender plays: the woman's odds ratio is $0.634$ (coefficient: -0.456, p-value: $p<0.001$, 95\% confidence interval of coefficient: $[-0.526,-0.385]$), indicating that the probability a woman artist enters the auction market is $0.634$ times that of a man.
Consider, for example, man artist A and woman artist B (Table ~\ref{tab:lr-case}), both with a career length of 11 years and 5 exhibits per year, and careers dominated by man-overrepresented institutions, i.e., they are both co-exhibition man artists.
The model predicts that the man artist A will transition to the auction market with probability $0.73$ while the woman artist B transitions only with probability $0.69$, reflecting the trend of man artists having more auction access.

In Model 3, we add co-exhibition gender, and found that the model BIC is further decreased.
Interestingly, the odds ratio of co-exhibition woman is $0.389$ (coefficient: -0.945, p-value: $p<0.001$, 95\% confidence interval of coefficient: $[-1.033,-0.856]$), indicating that the probability to auction an artist labelled as co-exhibition woman is $0.428$ times that of a co-exhibition man. This number is lower than the odds ratio for gender woman, which is $0.832$ (coefficient: -0.184, p-value: $p<0.001$, 95\% confidence interval of coefficient: $[-0.260,-0.109]$).
This difference in magnitude suggests that co-exhibition gender has a stronger correlation with the access to the auction market than the artists' gender.
To illustrate the effect of this difference, consider a comparison between artist A introduced above who is co-exhibition man and artist C, fully matched to artist A but with a co-exhibition woman designation (Table ~\ref{tab:lr-case}): while artist A has a $0.73$ probability of transitioning to the auction market, artist C has a transition probability of only $0.52$.
Overall, the odds ratios suggest that the gender-representation of institutions embracing an artist's work is a stronger indicator of the likelihood that an artist will transition to the auction market than the gender of the artist.

Finally, we explore the interaction of gender and co-exhibition gender in Model 4.  
This model has similar BIC as Model 3, indicating we gain little predictive power by increasing the model complexity.
However, the odds ratio for interactions between gender and co-exhibition gender are noteworthy.
A woman labelled as co-exhibition woman has an odds ratio of $0.321$ (coefficient: -1.137, p-value: $p<0.001$, 95\% confidence interval of coefficient: $[-1.24,-1.033]$) compared to a man labelled as co-exhibition man, yet this doubles to $0.754$ (coefficient: -0.283, p-value: $p<0.001$, 95\% confidence interval of coefficient: $[-0.42,-0.146]$) when the woman artist is labelled as co-exhibition man.
In other words, simply changing the community in which the artist associates is linked to a doubling of her chances of gaining access to the auction market.

Multiple model variations including the artists' prestige, nationality, and medium command similar patterns (Supplementary Information S2.4).
Taken together, our results indicate that the institutions which act as gatekeepers for the auction market show gender imbalance, and their gender representation inequality correlate with the valuation of the artists they embrace, independent of the gender of the artist.

\section*{Discussions}
In the 21st century, our society continues to struggle with gender imbalance and its implications---from business to science, men outnumber women, often dramatically~\cite{iversen2010women,stamarski2015gender,huang2020historical,holman2018gender}. 
Art is no exception, classical art (19th century and before) being represented by virtually all men, and contemporary art continues to struggle with gender-balance despite considerable efforts to acknowledge and correct it~\cite{topaz2019diversity,davis2016why}.

In this work, we explored institutional gender inequality using a comprehensive dataset of artist exhibitions and auctions. We find that 36.54\% of the contemporary, active artists are women, a number often quoted as the evidence of gender imbalance in art.
An additional gender disparity emerges if we inspect the opportunities offered to the practising artists, resulting in an exhibition gender ratio of 1.86, indicating that practising men artists are offered disproportionately more exhibition opportunities than practising women artists.
The gender disparity is particularly dramatic in the secondary auction market (Table~\ref{tab:bias_table}), indicating that art produced by men are seen as more valuable.

Controlling for the inherent population differences allowed us to quantify the extent to which individual institutions exhibited artists of a gender, showing overrepresentation of men artists for many (but not all) galleries and museums, especially for high prestige institutions.
Surprisingly, we find that artist's co-exhibition gender has a stronger correlation with the access to the secondary market than the artist's gender, suggesting that there might be relationship between institutional gender inequality and the auction market.

Our work opens directions for future research into potential interventions that may address the gender inequalities in the art world. For example, in education and academe, evidence-based training has been shown to reduce individual stereotyping and implicit bias~\cite{carnes2015breakbias, kossek2017opting}. 
Due to the documented network constraints in artistic careers, these interventions may have propagating influences beyond target institutions which would enhance their effectiveness. Our findings on gender-representation lock-in also suggest that potential interventions offering greater resources or opportunities to exhibit early in an artist’s career may have compounding influences. 

It is important to emphasize that these measures reflect emergent properties of the art world as an interconnected system, synthesizing the result of millions of local decisions by curators and museum and gallery administrators, made in the momentary best interest of their institutions. 
We think that awareness of these emerging outcomes are necessary for institutions to revisit their decision process, allowing them to rely on the proposed metrics to revise and guide their future policies.
Specifically, given the clustered structure of the artist's co-exhibition network (Fig.~\ref{fig:viz}, Supplementary Information S1.2), there may be interesting local patterns in institutional representation inequality which may be revealed by conditioning institutional representation on country-specific, temporal, community, or prestige relevant gender ratios of the artist population.
Reconstructing the co-exhibition network based on artist-nodes may facilitate more nuanced analysis of differences between galleries and museums on artists' careers, and would empower an ego network analysis of different exhibition patterns~\cite{lerman2022gendered}.
We want to emphasize that while our methods contribute towards a quantitative understanding of the gendered representational differences in art, they are not designed to identify the specific policies and decisions which create and perpetuate the barriers to representational equality.
Nonetheless, we believe our analysis has practical implications for museum administration, cultural policy, and philanthropy, offering a quantitative framework to diagnose the extent and the nature of the biases they face, and help them design appropriate policies to address them.

We must acknowledge some of the limitations of our work. 
First, while the exhibition data used in this study is extensive in both scale and coverage, it still offers an incomplete record, particularly when it comes to less prominent institutions and geographical regions (see data limitations discussed in Methods). Indeed, the observed gender representation inequality (or its lack of) for some institutions may be rooted in incomplete data. 
There is a need, therefore, for a community effort to reconstruct the exhibition history of all artists and institutions, independent of their perceived importance or prestige. Extending our data to include expert curated biographies could facilitate the application of more advanced gender identification tools~\cite{santamaria2018comparison}.
Additionally, we believe that the source of the institutional inequality cannot be fully understood without data on the curators and collectors, who are the driving force behind art collection, exhibition and auction, and often drive the behaviour of galleries. Such data, if it were to become available, would enable us to analyse the artistic ecosystem in a more holistic way.
Second, in this work, we assign most artists a binary gender using \href{https://genderize.io/}{genderize.io}, which has different coverage rate of names across ethnicity. 
Additionally, artists may affiliate across a spectrum of gender identities, prompting many artists to examine, question, and criticize the relationships between gender and society~\cite{moma}. While non-binary gender may also affect artistic success, we lack an aggregated database of artists to accurately infer non-binary gender information at scale. However, we believe that our work provides the statistical framework that can be combined with additional data to analyse inequalities impacting more fluid gender identifies.
Third, the present study focuses on the exhibitions of contemporary, active artists.
However, an equally important issue is the historical gender inequality, crystallized within institutional legacy collections and the consequential imbalanced exposure for museum visitors~\cite{topaz2019diversity}. 
Fourth, to ensure the robustness of our classifications, we excluded cases where the Bayes Factor fell within the ‘anecdotal evidence’ range ($1/3$ $<$ BF $<$ 3), as these values provide insufficient support for determining gender representation differences. 
While this approach reduces ambiguity, it results in some institutions remaining uncategorized due to insufficient statistical evidence.
Finally, while our analysis is limited to the signal associated with gender, there is a need for an equally important data-driven analysis on the role of ethnicity and non-binary gender categories, helping us unveil the institutional practices that affect the career of all minority participants in the art space.

We must also acknowledge some of the severe limitations of the gender-neutral approach: it ratifies and reinforces the structural inequities that created those gender imbalances to begin with. While gender-neutrality may seem a tempting and defensible institutional goal, it becomes a reflection of the inequities of society. Under the guise of fairness, it embeds discrimination into institutional practice, hence achieving gender-neutrality does not eliminate the potential need for affirmative action. Ultimately, gender equity in art would be reflected both in equal representation (equal populations, number of gallery and museum exhibitions, access to auctions, and auction sales), and gender-neutral movement and selection criteria such that artist careers are not additionally encumbered based on gender. The statistical and network-based frameworks that we introduce here address both concerns, revealing the levels of representational equality present in individual institutions, and allowing communities of institutions to jointly reflect on how inequality emerges from their collective behaviours.

Whatever conclusions are drawn from the diagnosis of an institution’s exhibition pattern, and whatever the institution defines as the desirable goal in terms of gender, the challenge will remain to establish institutional strategies and decision-making guidelines that will get all institutions to a more desirable place.
These strategies would need to be translated to transparent policies and efforts to successfully represent these in the court of public opinion. 
In other words, it is not sufficient in the long run to quantify the existing imbalances, raising awareness of their existence and extent, but such an analysis needs to be followed with steps to address the appropriate ethical, administrative, and political dimensions of the issue.
Taken together, our results suggest that to establish gender equality in art, it is not enough to only focus on enhancing the number of women artists, we must also break down the strong institutional representation inequalities which is related to artist mobility between institutions based on the artist's gender.

\section*{Methods}
In this section, we provide details of the methods in this work.
This research has obtained an exempt review from the Institutional Review Board (IRB) of Northeastern University with number IRB \#: 22-08-10.

\subsection*{Data} \label{sec:data}
\subsubsection*{Data Description}
Our dataset was collected by \href{https://artfacts.net}{Artfacts.net} (a leading data company operating in the art market) and pricing information was collected from \href{https://www.magnus.net}{magnus.net} (a company focused on pricing data in the art market). It combines information on artists' exhibits, auction sales, and primary market quotes, extending the information available on websites such as Artnet or Artprice.

To collect data on art exhibits, \href{https://artfacts.net}{Artfacts.net} ﬁrst identiﬁed institutions worldwide, using: i) online aggregators, such as \href{www.artforum.com}{www.artforum.com}, \href{www.index-berlin.de}{www.index-berlin.de}, and \href{www.chelseagallerymap.com}{www.chelseagallerymap.com}, ii) art institutions associations websites, like \href{www.artdealers.org}{www.artdealers.org}, and iii) art fairs websites, such as \href{www.artbasel.com}{www.artbasel.com}. For each institution, they then collected the information listed on their websites on all past exhibits. For each exhibit, they recorded the name of all exhibited artists, the date at which it occurred, and the geographic location where it took place. They also collected information on artists birth date, city of origin, gender as well as death date in case of non-living artists. The full exhibition dataset includes 787,473 art exhibitions at 23,570 institutions across 143 countries, and the disambiguated exhibition history for 496,354 artists. 
Note that in our work, we count exhibitions from the perspective of artist careers, such that if an exhibition involves multiple artists, we count it once for each artist equally (independent of the number of pieces/time exhibited).

Auction data is obtained by collecting all past transactions listed by the largest auction houses, such as Sotheby’s, Christie’s, or Phillips, as well as auction price aggregators. We have 3,257,886 auction records for the same time period. 
Each auction sale is characterized by its date, its realized price (converted to 2013 USD), the name of the artist, the name of the auction house and the artwork media. 
More information about data collection and validation can be found in Fraiberger et al., 2018 and their online Supplementary Information~\cite{fraiberger2018quantifying}.

\subsubsection*{Gender Identification}
The original data contain expert curated gender information for 207,392 artists (130,848 men and 76,544 women), covering about 41.78\% of the total population. 
To identify missing gender information, we use the tool \href{https://genderize.io}{genderize.io}, a publicly available service which identifies the likely gender based on first name. 
Using \href{https://genderize.io}{genderize.io}, we identified an additional 158,664 men artists and 105,741 women artists, giving a total of 471,791 artists with gender assignments (289,512 men artists and 182,285 women artists), covering about 93.04\% of the total population in the dataset.
For the analysis, we excluded 24,557 (4.95\%) artists for whom we could not infer a gender because the name had an ambiguous gender (probability $<$ 0.6) or was not included in the database.

\subsubsection*{Data Pre-processing}
Since our focus is on the evolution of artists' careers, for quality control, we filtered out those artists whose career started after age 50, or before age 18, resulting in 86,894 artists. 
The utility of this filter assumes that artists from antiquity would appear to start their careers late in life (i.e. after 50) because our data coverage before 1990 is sparse. 
We also filtered out exhibitions and auctions that occur before an artist's adulthood (age 18), suspecting data incorrectness for those records. 
Finally, we select contemporary artists whose careers start their exhibitions on and after 1990, resulting in our final 65,768 artists. 
After these filters, we end up with 65,768 artists participating in 1,018,077 exhibitions at 20,389 institutions worldwide and 125,682 auction records.

To account for the increasing price of auctions over time, we normalize each sale price by the annual average auction price, allowing us to compare artworks across artists and time. 

\subsubsection*{Data Limitations}
It is important to acknowledge several limitations to the dataset.
The \href{https://genderize.io}{genderize.io} gender identification algorithm employed here has a low coverage rate for Asian names, meaning that we may be unable to identify the gender of Asian artists, and thus our sample may underestimate artist participation from East Asian countries. 
Furthermore, first name based gender identification ignores the fact that gender is a spectrum.
Yet, we argue that this simplification allows us to capture large-scale statistical features of institutional gender bias in the art world.

Since the exhibition data is collected from each institution, it may reflect a biased coverage towards the more established institutions with more effort in digitization.
This may translate into incomplete artist careers if the artists exhibited at institutions not covered by the dataset, although, the missing institutions would likely be small and of lower prestige.
The data may also more accurately cover recent exhibitions due to the growing prevalence of digitized access to exhibition data.
Additionally, we notice that based on Figure \ref{fig:fig1}b and c, both the men and women curves peaked in 2006 for population and 2010 for exhibition.

There is a potential that our data is more well covered for the Western art world such as in United States and Germany. As a further check on the coverage of our data, we manually collected the artist exhibition data for all active Hungarian institutions~\cite{barabasi2020exhibit}.
Of the identified 640 Hungarian institutions, 82 appear in our dataset, resulting in a coverage rate of 12.81\%, indicating a relatively low coverage in smaller institutions. However, of those 82 institutions, we cover 75\% of the exhibition records. Additionally, those unmatched 558 institutions capture only 54\% of the exhibitions, indicating that the matched institutions are in generally large and prestigious institutions. Indeed, we see that among the top 20 largest institutions, we have a 70\% coverage rate.

\subsection*{Determine Institutional Representation Inequality}
To categorize each institution based on the representation of exhibited artists, we employ a binomial hypothesis test on the fraction of women exhibitions.
We first conduct the two-tail binomial test on $H_0$: $p= p_0$ (gender-neutral or gender-balanced) and $H_1$: $p \neq p_0$. 
We then calculate the Bayes Factor $BF_{10}^{1}$ for each hypothesis. 
If $BF_{10}^{1} < \frac{1}{3}$, we categorize the institution as gender-neutral or gender-balanced. 
If $\frac{1}{3} < BF_{10}^{1} <3$, it indicates that there is not sufficient evidence for either hypothesis, we exclude this institution for the analysis.

If $BF_{10}^{1} > 3$, we then determine whether an institution is women-overrepresented or men-overrepresented. 
Let's assume we have an institution where $p < p_0$, and we want to test whether we have enough evidence to confirm it's men-overrepresented. Then the two hypotheses are $H_0$: $p=p_0$ and $H_1$: $p<p_0$. We calculate Bayes Factor $BF_{10}^{2}$ for the above two hypotheses. If $BF_{10}^{2} > 3$, then we can categorize this institution as men-overrepresented. Otherwise, this institution is excluded from further analysis. We only need to flip the sign of $H_1$ for women-overrepresented.

\subsection*{Co-exhibition Gender Assignment}
The co-exhibition gender of an artist is assigned as the dominant institutional gender representation inequality of his/her exhibited institutions, based on the gender-neutral or gender-balanced criteria.
Specifically, we measure the percentage of type $\alpha$ institutions in an artist's exhibition career ($\rho_\alpha$), where  $\alpha\in$ \{{woman-underepresented}, {man-underepresented}, neutral or balanced\}.
We compare $\rho_\alpha$ against the expected percentage $\rho_\alpha^r$ of each type $\alpha$ of institution if it was selected at random.
We then calculate the relative difference between $\rho_\alpha$ and $\rho_\alpha^r$, and find the the category $\hat{\alpha}$ with the largest relative difference: 
\begin{equation}
\begin{aligned}
 \hat{\alpha} &= \text{argmax}_\alpha \frac{\rho_\alpha-\rho_\alpha^r}{\rho_\alpha^r},\\
 \alpha \in &\{\text{{woman-underepresented}}, \\
 &\ \text{{man-underepresented}}, \\
 &\ \text{neutral\ or\ balanced}\}.
\end{aligned}
\end{equation}

Then we assign co-exhibition gender to the artist based on $\hat{\alpha}$. If $\hat{\alpha}$ is {woman-underepresented}/{man-underepresented}/neutral or balanced, we consider this artist as co-exhibition man/co-exhibition woman/co-exhibition neutral or balanced.

In the main paper, we only showed results of co-exihibition gender categorised under the gender-neutral criteria. For results under gender-balanced criteria, see Supplementary Information S2.2.

\subsection*{Logistic Regression on Auction Access}
To build the logistic regression model to predict the artists' access to the auction market based on their exhibition history, the data was first pre-processed following these two steps:

\begin{itemize}
 \item For numerical features (number of exhibitions and the career length), we took the logarithm of the values and applied min-max normalization on the features. 
 \item For categorical features (such as gender, co-exhibition gender and prestige), we represented the values using dummy encoding, where the baselines are Man, Co-exhibition Man and High prestige, respectively.
\end{itemize}

The logistic regression model is then fitted on the processed data using Python package \texttt{statsmodels} \cite{seabold2010statsmodels}. For different variations of the logistic regression models see Supplementary Information Section S2.4.

\section*{Data Availability}
The data necessary to replicate the research are available at Harvard Dataverse under accession code 10.7910/DVN/ZTWCZR [\url{https://doi.org/10.7910/DVN/ZTWCZR}]~\cite{data}. 
Access to this dataset is restricted to noncommercial use and research purposes only.

\section*{Code Availability}
{The code to replicate the research can be found in the GitHub repository \url{https://github.com/Barabasi-Lab/artGender}~\cite{code}.}

\newpage

\singlespacing

\begingroup
\raggedright

\printbibliography

\section*{Acknowledgements}
We thank Andr\'{a}s Sz\'{a}nt\'{o}, Kathrin Zippel and Yasmine Ergas for insightful feedback on the manuscript, and Alice Grishchenko for help with the visualizations. We also thank the wonderful research community at the Center for Complex Network Research, particularly those in the success group, for valuable discussions and comments. X.W., A.J.G. and A.-L.B. were supported in part by Templeton Foundation Contract 61066 and Air Force Office of Scientific Research Award FA9550-19-1-0354; A.-L.B. is also supported by the European Union's Horizon 2020 research and innovation program under grant agreement No. 810115-DYNASNET, the Eric and Schmidt Fund for Strategic Innovation award \#G-22-63228, and NSF grant \#SES-2219575.

\section*{Author Contributions Statement}
{All authors discussed the results and commented on the manuscript. X.W analyzed the data, and developed the methods. X.W., A.J.G. and A-LB prepared the manuscript. M.R. provided the data. All authors read and approved the ﬁnal manuscript.}

\section*{Competing Interest Statement}
M.R. is CEO and Founder of Artace Inc, a company that collects and provides data for the art world.
The remaining authors declare no competing interests.

\newpage
\begin{table*}[]
 \centering
 \caption{Definition of Gender-Neutrality and Gender-Balance Hypotheses.}
 \resizebox{\textwidth}{!}{\begin{tabular}{m{3.2cm}m{4.1cm}m{6cm}}
  Name & Definition & Expected Women Representation\\ \hline\hline
  Gender-neutrality & artists have gender-independent, equitable access to representation & the fraction of exhibiting women artists (36.54\%)\\ \hline
  Gender-balance & 50/50 parity of men and women in population and representation & the fraction of women in the population (50\%) \\\hline
 \end{tabular}}
 \label{tab:def}
\end{table*}

\newpage
\begin{table}[]
\caption{Logistic Regression auction access prediction models \{showing Number of Observations, Degree of Freedom, Bayesian Information Criterion (BIC)\}, coefficients (coef.), odds ratios (O.R.), standard errors (S.E.), statistical significance (P val., {nominal, two-sided}) and  95\% confidence interval (Conf. Int.). Our confidence interval is calculated under $\alpha = 0.05$.}
\label{tab:lr}
\resizebox{\textwidth}{!}{%
\begin{tabular}{cccccccccccc}
\multicolumn{2}{c}{\textbf{Model}}        & \multicolumn{5}{c}{\textbf{(1)}}                              & \multicolumn{5}{c}{\textbf{(2)}}                              \\ \hline
\multicolumn{2}{c}{Number of Observations}& \multicolumn{10}{c}{16126}                                                                                                    \\ \hline
\multicolumn{2}{c}{Degree of Freedom}     & \multicolumn{5}{c}{2}                                         & \multicolumn{5}{c}{3}                                         \\ \hline
\multicolumn{2}{c}{BIC}                   & \multicolumn{5}{c}{20415.93}                                  & \multicolumn{5}{c}{20263.17}                                  \\ \hline
\multicolumn{2}{c}{Variables}             & Coef.  & O.R.   & S.E.  & P Val.           & Conf. Int.       & Coef.  & O.R.   & S.E.  & P Val.           & Conf. Int.       \\ \hline
\multicolumn{2}{c}{Intercept}             & -4.425 & 0.012  & 0.112 & \textless{}0.001 & (-4.645, -4.205) & -4.265 & 0.014  & 0.113 & \textless{}0.001 & (-4.486, -4.043) \\
\multicolumn{2}{c}{Exhibition Count Per Year}                                                                                                           & 3.794  & 44.435 & 0.129 & \textless{}0.001 & (3.542, 4.046)   & 3.809  & 45.087 & 0.129 & \textless{}0.001 & (3.555, 4.062)   \\
\multicolumn{2}{c}{Career Length}         & 4.233  & 68.914 & 0.127 & \textless{}0.001 & (3.984, 4.482)   & 4.213  & 67.561 & 0.128 & \textless{}0.001 & (3.963, 4.463)   \\
Gender (Baseline: Man)                                                                                                     & Woman                      & -      & -      & -     & -                & -                & -0.456 & 0.634  & 0.036 & \textless{}0.001 & (-0.526, -0.385) \\
\multirow{2}{*}{\begin{tabular}[c]{@{}c@{}}Co-exhibit. Gender \\ (Baseline: Co-exhibit. Man)\end{tabular}}                 & Co-exhibit. Neutral        & -      & -      & -     & -                & -                & -      & -      & -     & -                & -                \\
             & Co-exhibit. Woman          & -      & -      & -     & -                & -                & -      & -      & -     & -                & -                \\
\multirow{5}{*}{\begin{tabular}[c]{@{}c@{}}Gender x Co-exhibition Gender \\ (Baseline: Man, Co-exhibit. Man)\end{tabular}} & Woman, Co-exhibit. Neutral & -      & -      & -     & -                & -                & -      & -      & -     & -                & -                \\
             & Woman, Co-exhibit. Man     & -      & -      & -     & -                & -                & -      & -      & -     & -                & -                \\
             & Woman, Co-exhibit. Woman   & -      & -      & -     & -                & -                & -      & -      & -     & -                & -                \\
             & Man, Co-exhibit. Neutral   & -      & -      & -     & -                & -                & -      & -      & -     & -                & -                \\
             & Man, Co-exhibit. Woman     & -      & -      & -     & -                & -                & -      & -      & -     & -                & -                \\
\multicolumn{2}{c}{\textbf{Model}}        & \multicolumn{5}{c}{\textbf{(3)}}                              & \multicolumn{5}{c}{\textbf{(4)}}                              \\ \hline
\multicolumn{2}{c}{Number of Observations}& \multicolumn{10}{c}{16126}                                                                                                    \\ \hline
\multicolumn{2}{c}{Degree of Freedom}     & \multicolumn{5}{c}{5}                                         & \multicolumn{5}{c}{7}                                         \\ \hline
\multicolumn{2}{c}{BIC}                   & \multicolumn{5}{c}{19755.10}                                  & \multicolumn{5}{c}{19771.42}                                  \\ \hline
\multicolumn{2}{c}{Variables}             & Coef.  & O.R.   & S.E.  & P Val.           & Conf. Int.       & Coef.  & O.R.   & S.E.  & P Val.           & Conf. Int.       \\ \hline
\multicolumn{2}{c}{Intercept}             & -4.076 & 0.017  & 0.115 &                  & (-4.302, -3.851) & -4.053 & 0.017  & 0.116 & \textless{}0.001 & (-4.28, -3.826)  \\
\multicolumn{2}{c}{Exhibition Count Per Year}                                                                                                           & 3.929  & 50.848 & 0.133 & \textless{}0.001 & (3.669, 4.189)   & 3.918  & 50.322 & 0.133 & \textless{}0.001 & (3.658, 4.179)   \\
\multicolumn{2}{c}{Career Length}         & 4.437  & 84.531 & 0.13  & \textless{}0.001 & (4.181, 4.693)   & 4.434  & 84.258 & 0.131 & \textless{}0.001 & (4.178, 4.69)    \\
Gender (Baseline: Man)                                                                                                     & Woman                      & -0.184 & 0.832  & 0.039 & \textless{}0.001 & (-0.26, -0.109)  & -      & -      & -     & -                & -                \\
\multirow{2}{*}{\begin{tabular}[c]{@{}c@{}}Co-exhibit. Gender \\ (Baseline: Co-exhibit. Man)\end{tabular}}                 & Co-exhibit. Neutral        & -0.703 & 0.495  & 0.041 & \textless{}0.001 & (-0.784, -0.623) & -      & -      & -     & -                & -                \\
             & Co-exhibit. Woman          & -0.945 & 0.389  & 0.045 & \textless{}0.001 & (-1.033, -0.856) & -      & -      & -     & -                & -                \\
\multirow{5}{*}{\begin{tabular}[c]{@{}c@{}}Gender x Co-exhibition Gender \\ (Baseline: Man, Co-exhibit. Man)\end{tabular}} & Woman, Co-exhibit. Neutral & -      & -      & -     & -                & -                & -0.862 & 0.422  & 0.061 & \textless{}0.001 & (-0.982, -0.743) \\
             & Woman, Co-exhibit. Man     & -      & -      & -     & -                & -                & -0.283 & 0.754  & 0.07  & \textless{}0.001 & (-0.42, -0.146)  \\
             & Woman, Co-exhibit. Woman   & -      & -      & -     & -                & -                & -1.137 & 0.321  & 0.053 & \textless{}0.001 & (-1.24, -1.033)  \\
             & Man, Co-exhibit. Neutral   & -      & -      & -     & -                & -                & -0.741 & 0.477  & 0.048 & \textless{}0.001 & (-0.834, -0.648) \\
             & Man, Co-exhibit. Woman     & -      & -      & -     & -                & -                & -0.975 & 0.377  & 0.058 & \textless{}0.001 & (-1.088, -0.862)
\end{tabular}
}
\end{table}

\newpage
\begin{table}[]
\caption{Predicted auction access probability using logistic regression given the artists profile of three hypothetical artist A, B, C.}
\label{tab:lr-case}
\centering
\begin{tabular}{rccc}
Artist & A & B & C \\ \hline
Exhibition Count Per Year & 5 & 5 & 5 \\
Career Length & 11 & 11 & 11 \\
Gender & Man & Woman & Man \\
Co-exhibition Gender & Man & Man & Woman \\ \hline
Predicted Probability & 0.73 & 0.69 & 0.52
\end{tabular}%
\end{table}

\newpage
\begin{table*}[]
\caption{Metrics of Gender Bias in Art. The table lists the gender bias metrics explored in this study and their magnitude.}
\resizebox{\textwidth}{!}{\begin{tabular}[b]{ccccc}
 & Men& Women & Bias Name & \vbox{\hbox{\strut Gender Ratio}\hbox{\strut \ \ \ \ \ \ \ (M/W)}}\\
 \hline\hline
Artist Population& 41,738  & 24,030  & \small{Population Disparity} &  1.74\\
\hline
Number of Exhibitions & 662,571 & 355,506 & \small{Exhibition Disparity} & 1.86\\
\hline
Auction Population & 10,179 & 3,761 & \small{Auction Population Disparity} & 2.71\\
\hline
Auction Access Rate & 24.39\% & 15.65\% & \small{Auction Access Rate Disparity} & 1.56 \\
\hline
Number of Auction Records & 102,729 & 22,953 & \small{Auction Record Disparity} & 4.48\\
\hline
Average Auction Price & 0.7 & 0.4 & \small{Auction Price Disparity} & 1.75 \\
\hline
Average Number of Auctions & 10.1 & 6.1 & \small{Auction Count Disparity} & 1.66 \\
\hline
Auction Total Sales & $9.8\times 10^4$ & $1.3\times 10^4$ & \small{Auction Price Disparity} & 7.54 \\
\hline
\end{tabular}}
 \label{tab:bias_table}
\end{table*}

\newpage
\begin{figure}
 \centering
 \includegraphics[width=.7\textwidth]{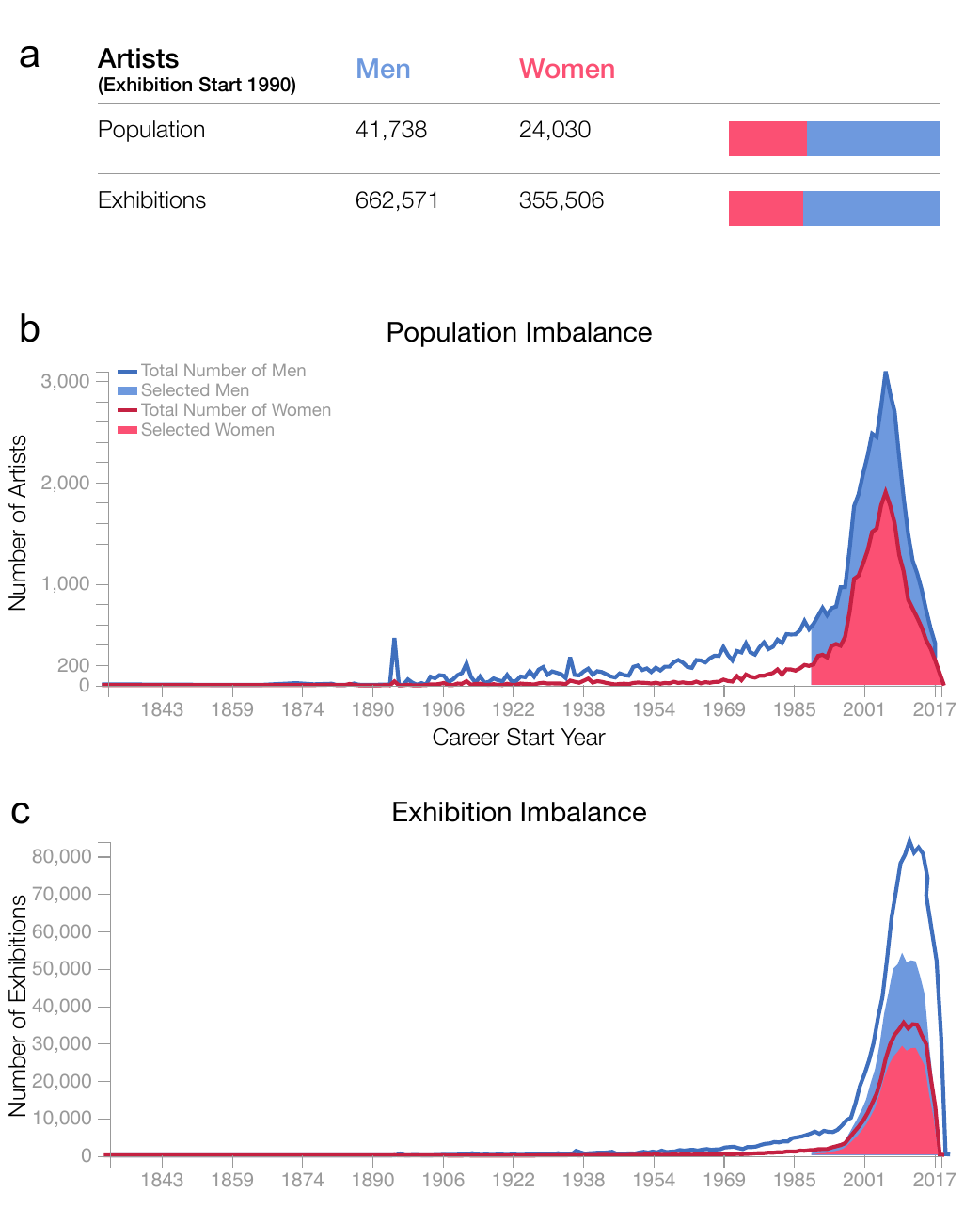}
 \caption{Gender Bias in Art. 
 (a) The gender-identified artist population and exhibitions studied in this paper. See Methods section for the data cleaning and filtering process that resulted in the explored artist pool.
 (b) Number of artists over time, showing separately the number of men and women artists. The shadowed area corresponds to artists and exhibitions selected for our study.
 (c) Number of exhibitions over time, shown separately for men and women artists. The shadowed area corresponds to artists and exhibitions selected for our study.
 }
 \label{fig:fig1}
\end{figure}

\begin{figure}
 \centering
 \includegraphics[width=.5\textwidth]{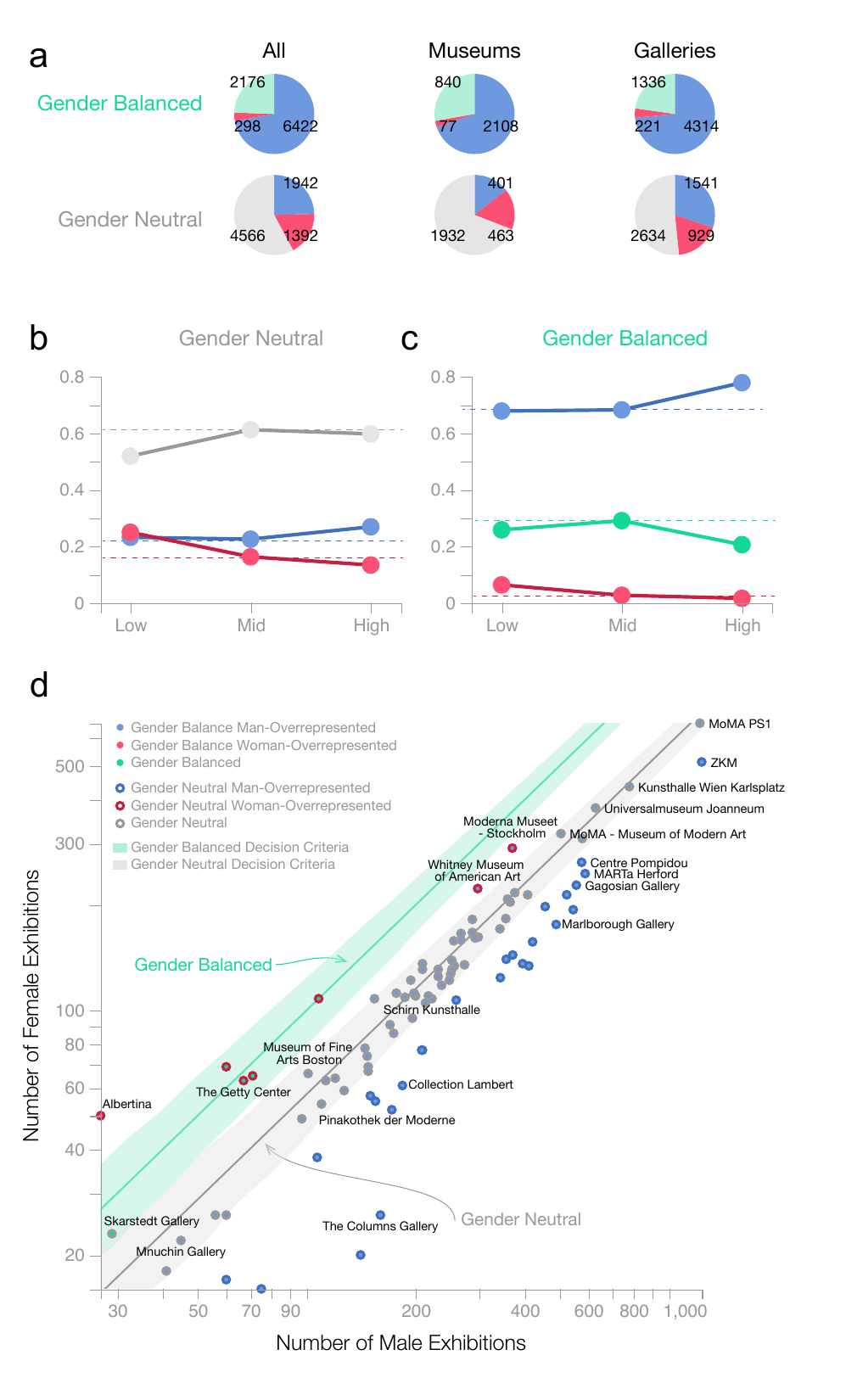}
 \caption{Gender-Balanced vs Gender-Neutral Institutional Patterns.
(a) Number of {man-overrepresented}, {woman-overrepresented} and gender-neutral or gender-balanced institutions under the gender-neutral and the gender-balanced criteria.
(b) Portion of {man-overrepresented}, {woman-overrepresented}, gender-neutral institutions of different institution prestige under the gender-neutral criteria.
(c) Portion of {man-overrepresented}, {woman-overrepresented}, gender-balanced institutions of different institution prestige under the gender-balanced criteria.
(d) Number of men and women exhibitions at 100 prominent institutions. The example compares an institution's status to the decision area under both gender-neutral and gender-balanced criteria. For example, MoMA (The Museum of Modern Art) is a {woman-overrepresented} institution under the gender-neutral criteria but {man-overrepresented} under the gender-balanced criteria.}
 \label{fig:fig2}
\end{figure}

\begin{figure}
 \centering
 \includegraphics[width=.6\textwidth]{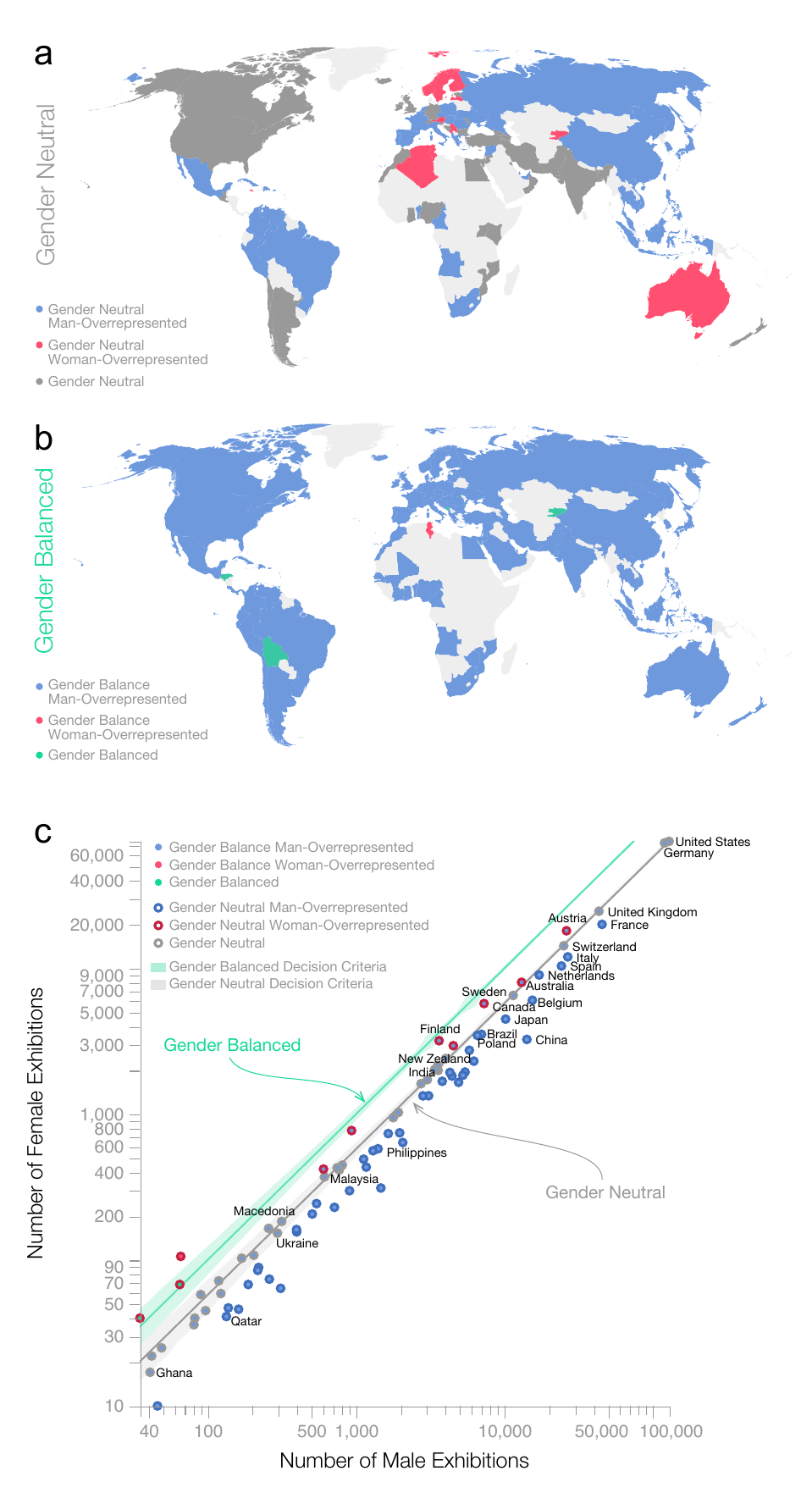}
 \caption{Institutional Gender Representation Inequality Across the World.
World map showing for each country its gender bias in art under the (a) gender-neutral and under the (b) gender-balanced criteria.
(c) Number of men and women exhibitions of countries/regions. The gray shading shows the gender-neutral decision area and green shading shows gender-balanced decision area. For example, the figure reveals that Finland is a {woman-overrepresented} country under the gender-neutral criteria but {man-overrepresented} under the gender-balanced criteria.}
\label{fig:fig3}
\end{figure}

\begin{figure}
 \centering
 \includegraphics[width=.6\textwidth]{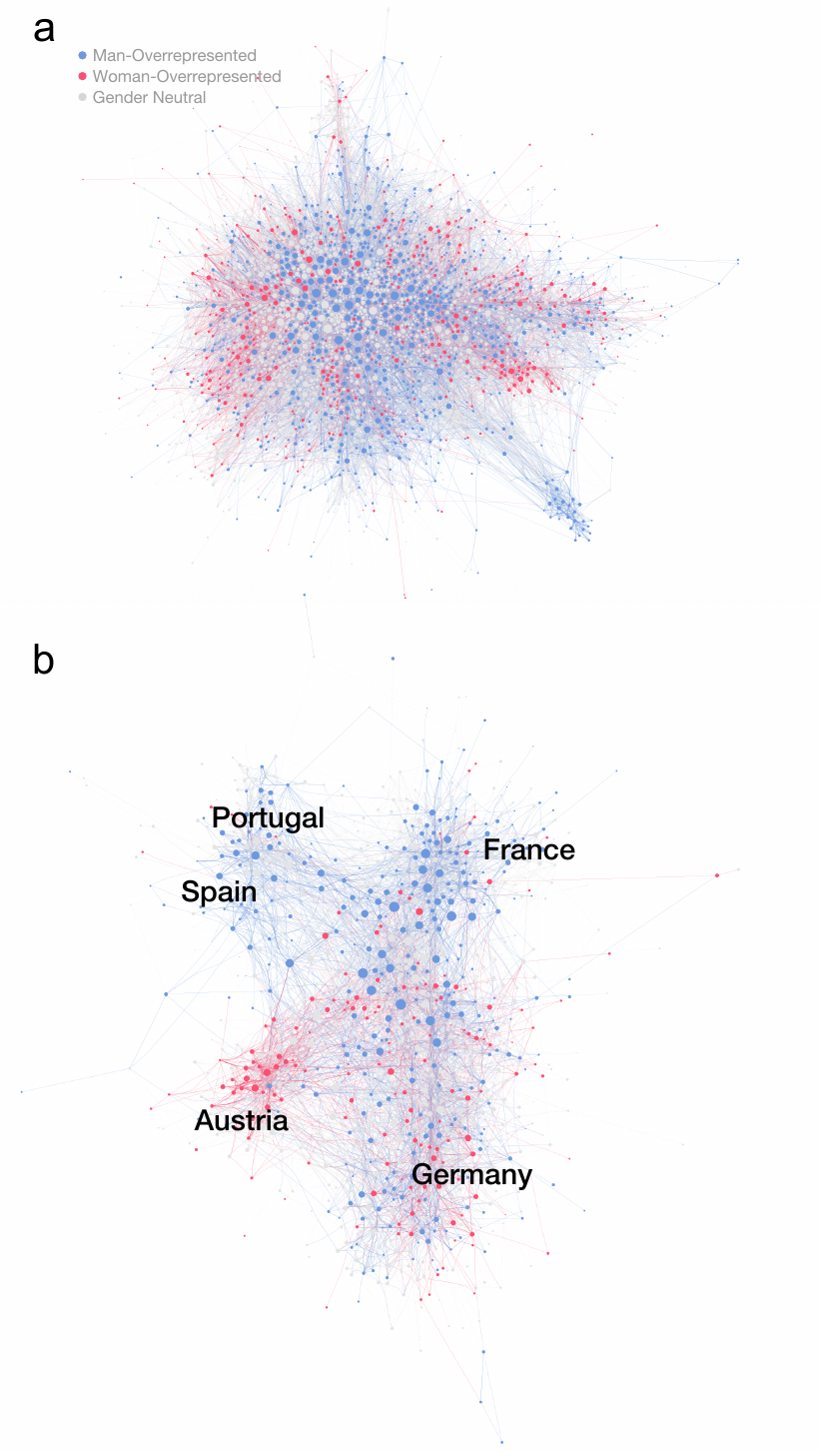}
 \caption{Institutional Co-exhibition Network. (a) Each node is an art institution coloured by the institutional gender representation under the gender-neutral criteria: {man-overrepresented}, {woman-overrepresented} or gender-neutral. Two institutions are connected by a directed, weighted link reflecting the number of artist-exhibitions in which the artist first exhibited in the source institution before later exhibiting in the target institution. The figure illustrates the inequality-based assortativity, namely, institutions with the same inequality status are more likely to connect to each other and belong to the same cluster. To better illustrate this effect, in (b) we show the sub-graph of several European countries: Germany, Austria, France, Spain and Portugal, allowing us to see the regional differences: Germany and Austria have more {woman-overrepresented} institutions while France, Spain and Portugal have more {man-overrepresented} institutions. We can also observe the presence of clusters based on gender representation within a region: {woman-overrepresented} institutions in Austria are close to {woman-overrepresented} institutions in Germany, while the few {man-overrepresented} institutions in Austria are closer to the other countries. For more details about network assortativity see Supplementary Information S1.2.}
\label{fig:viz}
\end{figure}

\begin{figure}
 \centering
 \includegraphics[width=.7\textwidth]{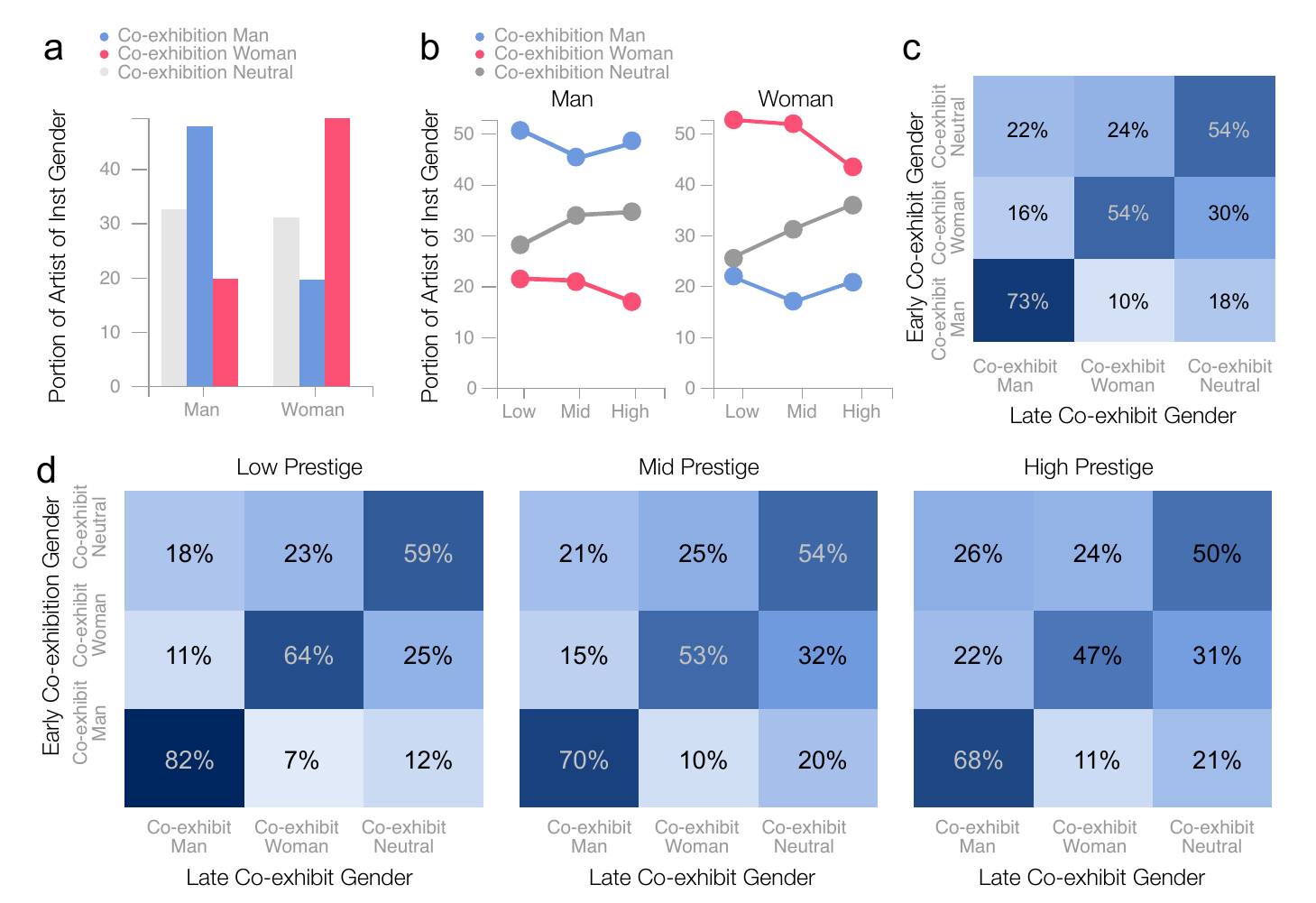}
 \caption{Co-exhibition Gender of Artist under Gender-Neutral Criteria.
 (a) Fraction of co-exhibition man, co-exhibition woman, and co-exhibition neutral for men and women artists based on the gender-neutral criteria.
 (b) Fraction of co-exhibition gender for men and women artists based on the gender-neutral criteria, with different career prestige.
 (c) Transition probability of early co-exhibition gender to late co-exhibition gender based on the gender-neutral criteria.
 (d) Transition probability of early co-exhibition gender to late co-exhibition gender based on the gender-neutral criteria for artists with different career prestige.}
 \label{fig:fig5}
\end{figure}

\begin{figure*}
 \centering
 \includegraphics[width=.7\textwidth]{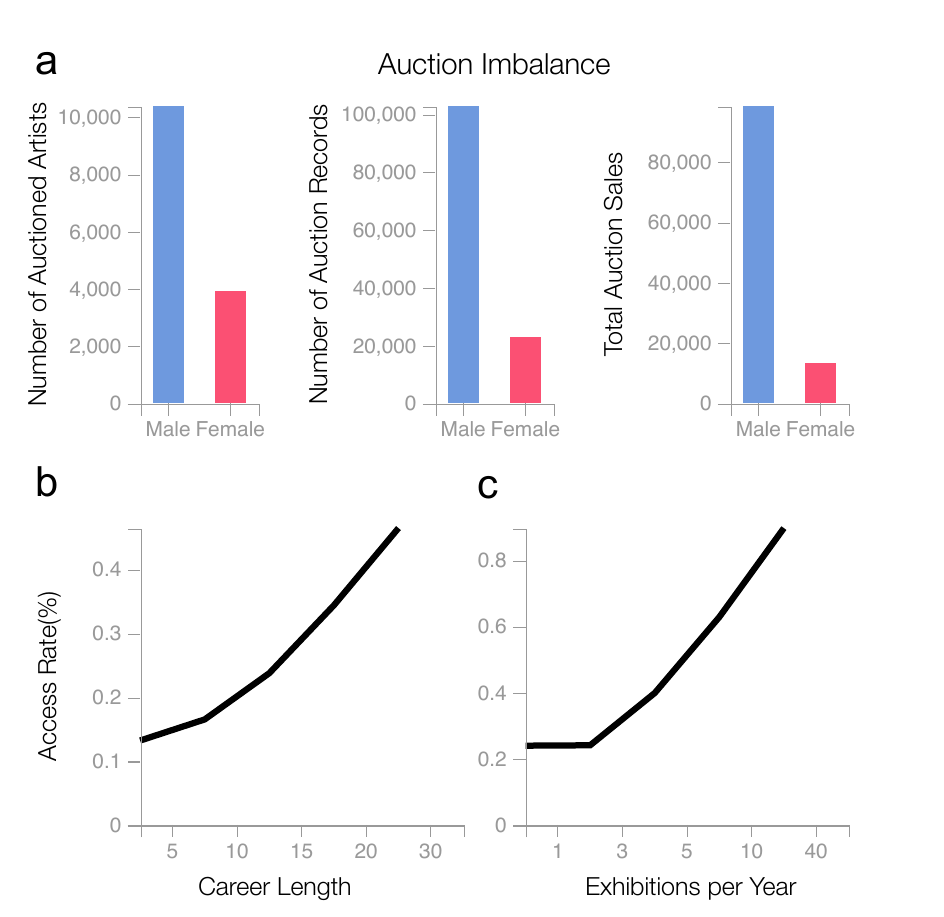}
 \caption{Gender Bias in Auction.
 (a) Gender bias in terms of auction population, number of auction records and total auction sales.
 (b) Access rate vs career length, indicating that the longer an artist's career, higher the chances that of the artist enters auctions.
 (c) Access rate vs exhibition count per year, indicating that the higher the exhibition count per year, higher the chances that of the artist enters auctions.
 }
 \label{fig:fig6}
\end{figure*}

\end{document}